# Do we understand the emergent dynamics of grid cell activity?

Yoram Burak and Ila Fiete

Kavli Institute for Theoretical Physics, University of California, Santa Barbara 93106

Single neurons in the dorsolateral band of the rat entorhinal cortex (dMEC) fire as a function of rat position whenever the rat is on any vertex of a regular triangular lattice, that tiles the entire plane (Hafting et al., 2005). Even in the dark, the pattern refreshes correctly with rat movement, and maintains its coherence over paths whose accumulated length exceeds 200 meters (Hafting et al., 2005). For dMEC activity to be a precise function of the rat's changing position in the dark, it must have access to an accurate estimate of position based on self-motion cues, suggesting that dMEC may be involved in idiothetic path integration.

**Model:** How can these remarkable properties emerge from simple cortical connectivity and network dynamics? To answer this question, Fuhs and Touretzky, in their recent Journal of Neuroscience paper (LINK), propose a model based on two main ingredients. The first ingredient is the formation of a static, hexagonal firing pattern within a 2-dimensional neural sheet. Fuhs & Touretzky suggest a specific topographic synaptic connectivity to show how such a pattern could form (Fuhs & Touretzky, 2006, Figures 1E, 2A). Indeed, triangular lattice patterns emerge quite generically and robustly in networks with local excitatory and longer-range local inhibitory connectivity that is symmetric and translation-invariant (Murray, 2003).

The *static* hexagonal pattern within the neural sheet does not yet explain the experimental data, in which individual neurons dynamically start or stop firing in response to changes in rat position (Figure 1A). The role of the second model ingredient is to explain the single neuron measurements, by establishing a link between rat movements in real space and pattern dynamics in the neural sheet. Hypothetically, if rat velocity inputs could induce pure translations of the firing pattern on the neural sheet, in exact proportion to the rat's translations in space, then as illustrated in Figure 1B, the model single neuron responses would resemble the experimentally observed grid cell activity. Moreover, nearby cells on the neural sheet would have different phases as a function of position, but would share the same period and orientation.

To achieve the above goal, Fuhs & Touretzky propose that specific neurons in dMEC receive input from specific head direction cells, and thus are preferentially excited when the rat is moving in a particular direction. In addition, the outgoing weights of these neurons are slightly biased in a matching direction within the neural sheet. When, due to rightward motion by the rat, rightward biased neurons receive larger velocity input than leftward biased neurons, they drive a corresponding rightward flow of the network pattern (Fuhs and Touretzky, 2006, Figure 2A). Similar machinery has been suggested or used to translate activity patterns in other models of neural integration (Zhang, 1996; Xie et al., 2002).

**Critical evaluation:** In their model, Fuhs & Touretzky demonstrate that a *brief* (300 ms) velocity input can move the population lattice pattern in the correct direction (Fuhs and Touretzky, 2006, Figure 2A). But they don't demonstrate or test whether their model can reproduce the *central experimental finding* of Hafting et al: that single neuron firing over a long (~200 m), variable speed rat trajectory is grid-like.

Here we generate single-neuron responses from the model when the network is subject to velocity inputs that mimic a rat moving around a 1 $m^2$ enclosure. The single-neuron response we obtain from Fuhs and Touretzky's model, unlike the experimental data, has no coherent grid structure (Figure 1C).

If the network pattern translates in the correct direction for brief velocity inputs, why does it not reproduce the single neuron responses? To produce single-neuron grids, the network pattern must flow exactly in register with rat position: in other words, the network must precisely integrate rat velocity. The lack of coherent single-neuron grids implies that the network does not integrate correctly, as confirmed by the tracking data of Figure 1D. By dissecting the network dynamics during the realistic trajectory of Figure

1C and during constant velocity steps, we observe that the network's velocity response is neither purely translational (Figure 1E), nor linear (Figure 1F).

Beyond the generic sensitivities of previously studied continuous attractor models of integration (Seung, 1996; Zhang, 1996; Xie et al., 2002), we can identify two reasons, specific to the symmetry and form of the triangular lattice network, for why it is difficult to obtain network dynamics that reproduce grid cell firing.

First, the full rotational symmetry of both the network boundary and the synaptic connections cause the continuous attractor of the network dynamics to contain all rotations, in addition to all translations, of the lattice pattern. Most of these rotations are distinct network states, and from the point of view of integration -- which requires purely translational responses to velocity inputs -- are undesirable.

Second, the network's aperiodic boundaries[1] serve to make the network dynamics state-dependent, instead of only input-dependent. We find in our own work (Burak and Fiete, unpublished observations) on similar attractor models of the grid cell system that this state-dependence can profoundly affect the linearity of the network velocity response and how often undesired rotated attractor states are reached. Boundary effects are more prominent in smaller networks, and are pronounced in networks small enough to be consistent with neuron number estimates in the entorhinal cortex.

These effects occur in fully deterministic idealized networks, and will likely be exacerbated by stochasticity in neural firing or non-uniformity in network connectivity.

---

[1] Periodic boundary conditions would require the physical connectivity in the neural sheet to have a torus-like topology. Such connectivity seems unrealistic for most networks of the brain, except perhaps in exceptional cases like the head direction network, which codes for a 1-dimensional variable that is itself periodic, and may therefore conceivably use an activity-dependent plasticity mechanism to generate periodic connectivity.

**Discussion:** The model of Fuhs & Touretzky is based on continuous attractor dynamics (Seung, 1996).[2] Conceptually, it accounts for many interesting properties of grid cells, such as the shared orientation but distributed phases of grid cells with the same periodicity. It also provides a conceptual explanation of how an endlessly repeating pattern that tiles enclosures of any size could arise from a finite population of neurons, without imposing periodic boundary conditions on the cortical sheet, Figure 1G.

However, the specific model of Fuhs and Touretzky is unable to reproduce the basic experimental observation, that the activity of single neurons in dMEC, plotted against rat position over long paths, forms a triangular lattice (Hafting et al., 2005). Can continuous attractor models, with structural or parameter adjustments, reproduce the grid cell data within known biological constraints on network size, neural noise, grid periods, and rat velocities? Or is the present model's lack of grid coherency symptomatic of fundamental limits that would necessitate the search for qualitatively new models of grid cell dynamics?

We note that the coherency of single-neuron grids over long paths in the dark is a remarkably stringent assay of the accuracy of neural position estimation from idiothetic cues: the accumulated error in the network's velocity-based flow over the full path must be small compared not merely to the path length (~200 m), but to a single grid period (~0.5 m). Even without knowing whether the rat behaviorally displays path integration over long journeys, we *can* unequivocally deduce from the single-neuron data that such information, albeit in modulo form, is present with striking accuracy in dMEC.

---

[2] For this reason, the model label "spin glass" is, in our opinion, inappropriate. Spin glasses are characterized by a landscape with multiple inequivalent local minima, separated by energetic barriers. The present model is a *continuous attractor model,* with a very different landscape containing a continuum of equivalent minima or attractors unseparated by barriers.

Quantitative results from experiment will be instrumental in helping to develop a mechanistic understanding of grid cell dynamics, and in ruling in or out a continuous attractor model: What is the path length scale for decoherence of grid responses in the dark, and is it a function of rat speed, or acceleration, or time? Can grid cell orientation rotate freely, relative to head direction cells, or is it pinned?

In summary, explaining the remarkable coherency of single-neuron dMEC grid patterns poses a central theoretical problem. The model of Fuhs & Touretzky provides a simple and appealing framework in which this question can be probed. Its plausibility as a *qualitative* model may hinge on its *quantitative* properties: whether it can produce the observed responses with realistic assumptions, and predict the signature vulnerabilities of the biological system.

**FIGURE CAPTION**

A. Synaptic connectivity with local excitation and longer-range local inhibition (not shown) leads to the formation of triangular lattice patterns on the neural sheet. Right: uncoupled to velocity inputs, the pattern is static: individual neurons are always on or always off, independent of rat position.

B. Single neuron responses would resemble dMEC activity if the network pattern performed rigid translations in direct proportion to rat velocity. The figure is generated by artificially translating the population activity in tandem with simulated rat movements, and plotting the response of a representative model neuron from the neural sheet.

C. Single-neuron responses (left panels), with their autocorrelations (right) generated from the Fuhs & Touretzky model using a semi-random rat trajectory (over 4 minutes intervals, ~120 meters), do not resemble dMEC activity. Simulation details are from (Fuhs and Touretzky, 2006), but we modified the narrow velocity tuning curves of (Fuhs and Touretzky, 2006) into cosine tuning curves to provide equitably weighted velocity input to the network for motion along any direction.

D. Rat trajectory (black) and tracked network response (blue) over ~50 m of movement at speeds < 0.5 m/sec (only x-coordinate shown, y is similar). Red scale bar: 1 lattice period, compared with which errors should be small for coherent single-neuron grids.

E. The network (population) pattern undergoes frequent rotations during the trajectory used in C, D above, instead of performing pure translation.

F. Network response, $v_{net}$, to velocity inputs, $v_{in}$: instead of varying linearly as a function of only rat velocity, network response is pinned for rat speeds < ~15 cm/sec (when the network gain is set to produce single neuron activities of period ~35 cm), and depends on the relative angle between lattice orientation and rat velocity. The network received 5-sec steps of constant-velocity input in the x direction, and the pattern's translation was measured over the last 2.5 sec. Red diamonds and blue circles correspond to two different initial pattern orientations relative to the x axis.

G. Even though the neural network boundary is aperiodic, the network dynamics are periodic by virtue of the periodicity of the neural pattern and due to spontaneous pattern completion during the velocity-driven flow of network activity.

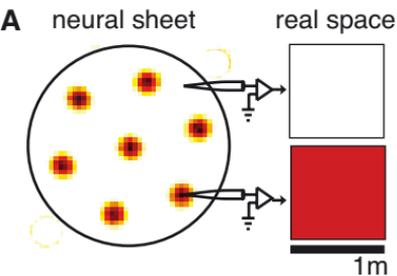
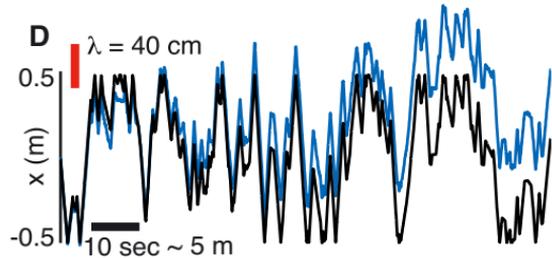
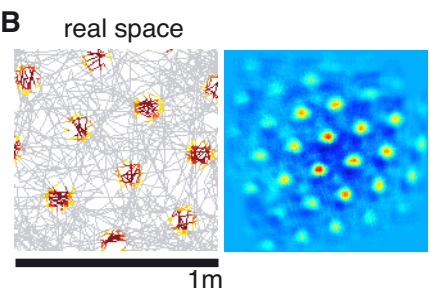
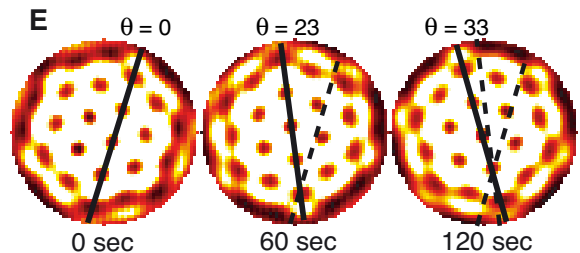
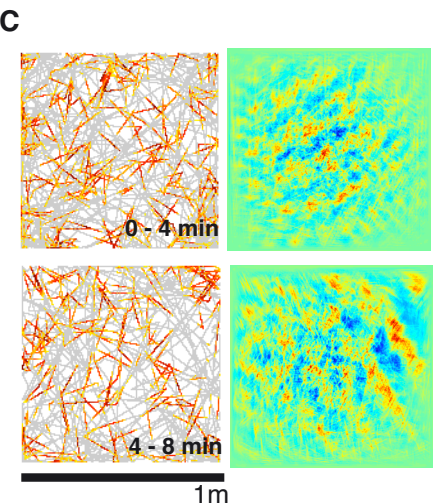
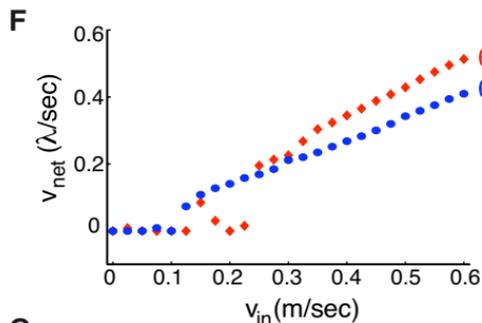
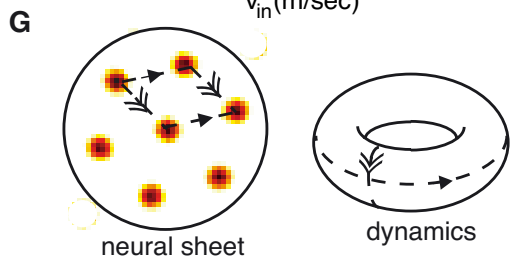